\documentclass[aps,superscriptaddress,pra,twocolumn,showpacs,accepted=2018-08-27]{quantumarticle}
\usepackage{amsfonts, amsmath, amssymb, bm, bbm, graphicx, epsfig, epstopdf}
\usepackage{braket}
\usepackage{dcolumn}
\usepackage{textcomp}
\usepackage{hyperref}
\usepackage{color}
\definecolor{forestgreen}{RGB}{34,139,34}

\usepackage[normalem]{ulem} 
\usepackage{color} 

\begin{document}
	\title{Quantum repeaters with individual rare-earth ions at telecommunication wavelengths}
	
\author{F. Kimiaee Asadi}
\affiliation{Institute for Quantum Science and Technology, and Department of Physics \& Astronomy, University of Calgary, 2500 University Drive NW, Calgary, Alberta T2N 1N4, Canada}

\author{N. Lauk}
\affiliation{Institute for Quantum Science and Technology, and Department of Physics \& Astronomy, University of Calgary, 2500 University Drive NW, Calgary, Alberta T2N 1N4, Canada}
\author{S. Wein}
\affiliation{Institute for Quantum Science and Technology, and Department of Physics \& Astronomy, University of Calgary, 2500 University Drive NW, Calgary, Alberta T2N 1N4, Canada}

\author{N. Sinclair}
\affiliation{Institute for Quantum Science and Technology, and Department of Physics \& Astronomy, University of Calgary, 2500 University Drive NW, Calgary, Alberta T2N 1N4, Canada}

\author{C. O'Brien}
\affiliation{Institute for Quantum Science and Technology, and Department of Physics \& Astronomy, University of Calgary, 2500 University Drive NW, Calgary, Alberta T2N 1N4, Canada}
\author{C. Simon}
\affiliation{Institute for Quantum Science and Technology, and Department of Physics \& Astronomy, University of Calgary, 2500 University Drive NW, Calgary, Alberta T2N 1N4, Canada}
	
	\begin{abstract}
        We present a quantum repeater scheme that is based on individual erbium and europium ions. erbium ions are attractive because they emit photons at telecommunication wavelength, while europium ions offer exceptional spin coherence for long-term storage. Entanglement between distant erbium ions is created by photon detection. The photon emission rate of each erbium ion is enhanced by a microcavity with high Purcell factor, as has recently been demonstrated. Entanglement is then transferred to nearby europium ions for storage. Gate operations between nearby ions are performed using dynamically controlled electric-dipole coupling. These gate operations allow entanglement swapping to be employed in order to extend the distance over which entanglement is distributed. The deterministic character of the gate operations allows improved entanglement distribution rates in comparison to atomic ensemble-based protocols. We also propose an approach that utilizes multiplexing in order to enhance the entanglement distribution rate. 
	\end{abstract}
	\maketitle
	\section{Introduction}
		Entanglement is a key requirement for many applications of quantum science. These include, for example, quantum key distribution \cite{QKD}, global clock networks \cite{clock}, long-baseline telescopes \cite{telescope}, and the quantum internet \cite{kimble, Christoph}. However, due to transmission loss the direct transmission of entanglement over distances of more than several hundred kilometers is practically impossible using current technology. The use of a quantum repeater has been suggested to reduce (or eliminate) the impact of loss in order to establish entanglement between distant locations. \cite{Briegel}. In many quantum repeater schemes, entanglement is first distributed between two locations that are separated by a short distance, referred to as an elementary link. Then, the range of entanglement is extended to successively longer distances by performing entanglement swapping operations between the entangled states that span each elementary link. Due to the availability and diversity of component systems and the strong light-matter coupling offered by atomic ensembles, many quantum repeater proposals use sources of entanglement, ensemble-based quantum memories, and linear optics-based entanglement swapping operations \cite{review-multi1}. However, the success probability of linear optics-based entanglement swapping (without auxiliary photons) cannot exceed 50\%, which has a compounding effect for more complex quantum networks. The use of auxiliary photons to improve the entanglement swapping probability is possible \cite{grice2011,wein2016}, but adds complexity and compounds errors, thereby restricting their use in practice.  Single-emitter-based quantum repeaters, on the other hand, offer the possibility to outperform ensemble-based repeaters by using deterministic swapping operations \cite{singletrappedions}. Impressive demonstrations of certain parts of a single-emitter quantum repeater schemes have been performed using atom-cavity systems \cite{Ritter2012, Reiserer2015}, color centers in diamond \cite{Hanson, Hensen2015}, trapped ions \cite{Moehring2007, Slodicka2013}, donor qubits in silicon \cite{silicon}, as well as quantum dots \cite{Delteil2016, Stockill2017}.
	
	Over the years, crystals doped with rare earth (RE) ions have attracted considerable attention for their use in electromagnetic signal processing applications that range from quantum memories \cite{tittelreview,bussieres} to biological imaging \cite{bio}. Narrow optical and spin homogeneous linewidths, their convenient wavelengths as well as their ability to be doped into solid-state crystals are some of the most desired properties of RE ions. Compared to nitrogen-vacancy centers in diamond \cite{diamond} and quantum dots \cite{dot}, they also exhibit smaller spectral diffusion \cite{spectral1,spectral2,wratch}, and have a reduced sensitivity to phonons.
	
	Among the different RE ions, Er$^{3+}$ is attractive due to its well-known optical transition (around 1.5 $\mu$m) in the conventional telecommunication wavelength window, in which absorption losses in optical fibers are minimal. Another unique aspect of certain RE ions is the presence of hyperfine levels that feature long lifetimes, which allows for long-term quantum state storage \cite{quantummemory1,quantummemory2, QIP2, afc}. In particular, a coherence lifetime of six hours was reported in a europium-doped yttrium orthosilicate crystal ($^{151}$Eu$^{3+}$:Y$_2$SiO$_5$) \cite{6hours}. Motivated by these properties, we propose and analyze a quantum repeater protocol that is based on individual RE ions in which  Er$^{3+}$ ions are used to establish entanglement over elementary links and $^{151}$Eu$^{3+}$ ions are employed to store this entanglement.
	
	One disadvantage of RE ions is their weak light-matter coupling, which has mostly precluded their use as single quantum emitters. However, optical detection and addressing of single RE ions has recently been shown by multiple groups \cite{ zhong2018, Thompson, Kolesov2012, Sellars2013, Eichhammer2015}. Moreover, very recently Dibos et al. \cite{Thompson} demonstrated an enhancement in the emission rate of a single Er$^{3+}$ ion in Y$_2$SiO$_5$ by a factor of more that 300 using a silicon nanophotonic cavity. Strong coupling of ensembles of Nd ions was previously demonstrated using nanophotonic cavities fabricated from Y$_2$SiO$_5$ \cite{faraon} and yttrium orthovanadate \cite{faraon2} hosts. In light of these results, for our scheme we propose to couple single RE ions to a high-finesse cavity in order to enhance the light-matter coupling and thus to increase the collection efficiency as well as the indistinguishability of the emitted single photons.
	 
	The paper is organized as follows. In Sec. \ref{Sec:Proposal} we introduce our proposal and discuss the required components as well as the underlying principles. The entanglement distribution rates and possible implementations are discussed in Secs. \ref{Sec:rate} and \ref{Sec:Implementation}. We conclude and provide an outlook in Sec. \ref{Sec:Conclusions}.
	
	\section{Quantum repeater protocol}\label{Sec:Proposal}
	The goal of our repeater scheme is to generate entanglement between nodes that are separated by a long distance  by swapping entanglement that is established between nodes that are separated by smaller distances. Each node consists of a Y$_2$SiO$_5$ crystal containing an Er$^{3+}$ ion that features an enhanced emission rate of its telecommunication transition due to a high-finesse cavity, and a nearby $^{151}$Eu$^{3+}$ ion that acts as a quantum memory due to its long spin coherence lifetime. Entanglement is transferred, or gates are performed, between nearby atoms using electric dipole-dipole coupling \cite{longdell1,longdell2}. Our proposal has some similarity to a scheme that involves the excitation of electron spins in nitrogen-vacancy centers to generate entanglement between nodes, nuclear spins of carbon atoms for storage, and gates that are performed using a magnetic dipole-dipole coupling \cite{NV-Carbon}. We emphasize that our (electric dipole) approach to implementing gates is quite different than previous approaches, as are many of the physical properties of the system (in addition to the emission wavelength).
    
\begin{figure}[!h]
		\includegraphics[width=8cm]{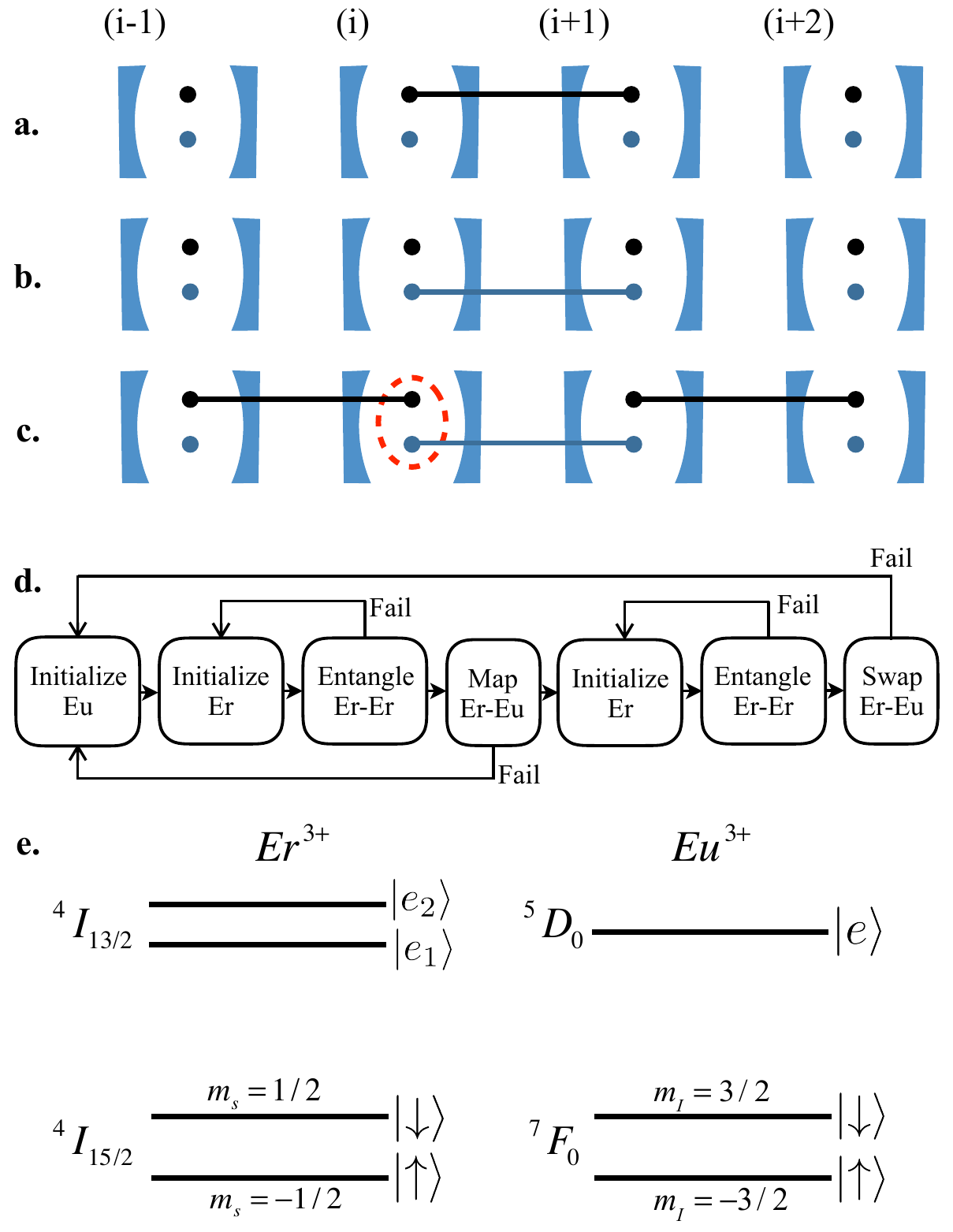}
		\caption{In each cavity, the blue and black dots represent individual $^{151}$Eu$^{3+}$ and Er$^{3+}$ ions, respectively. In order to distribute entanglement to the nodes that are situated at each end of a long channel, first a) entanglement between two neighboring Er$^{3+}$ ions is created via the detection of two photons; b) the quantum state of each Er$^{3+}$ ion is then mapped to a nearby $^{151}$Eu$^{3+}$ ion using an electric dipole interaction; c) using two-photon detection, Er$^{3+}$ ions are used again to establish entanglement between the other neighboring Er$^{3+}$ ions. Finally, we extend the entanglement distance by performing entanglement swapping using an electric dipole interaction between nearby ions (red circle). d) Flow-chart that depicts our scheme. e) Simplified energy level structure of Er$^{3+}$ and $^{151}$Eu$^{3+}$ ions in the presence of an external magnetic field. The electronic Zeeman levels of Er$^{3+}$, and nuclear hyperfine levels of $^{151}$Eu$^{3+}$, are split. We encode qubit states $\ket{\uparrow}$ and $\ket{\downarrow}$ in the $m_s=-\frac{1}{2}$ and $m_s=\frac{1}{2}$ $\left(m_I=-\frac{3}{2} \text{ and } m_I=\frac{3}{2}\right)$ levels for the Er$^{3+}$  ($^{151}$Eu$^{3+}$) ion, respectively.}\label{fig:main}
	\end{figure}

	In the first step, entanglement between pairs of neighboring Er$^{3+}$ ions is created (see Fig.\ref{fig:main} (a)) which illustrates this step for the i-th and i+1-th ions) by first creating local entanglement between the spin state of each ion and a spontaneously-emitted single photon. Then, by performing a joint Bell-state measurement (BSM) on photons that are spontaneously-emitted by neighboring Er$^{3+}$ ions, the spin state of these ions is projected onto a maximally-entangled state. This procedure is the same as that used to create entanglement between two remote nitrogen-vacancy centers in diamond \cite{barrett,Hanson}. 
	
	Next, the quantum state of each Er$^{3+}$ ion is mapped to a nearby $^{151}$Eu$^{3+}$ ion for storage by exploiting a non-vanishing permanent electric dipole moment, a common feature of many RE ions that are doped into solids. This mapping is achieved using a mutual electric dipole-dipole interaction between close-lying Er$^{3+}$ and $^{151}$Eu$^{3+}$ ions. The small nuclear magnetic moment of $^{151}$Eu$^{3+}$ results in a magnetic-dipole coupling that is orders of magnitude smaller than the electric-dipole coupling, thus our scheme allows for much shorter gate durations in comparison to those based on magnetic interactions \cite{NV-Carbon}. Another advantage of the proposed scheme is the ability to dynamically control the interaction optically by bringing the ion to the excited state and back to the ground state. This allows for the realization of deterministic two-qubit gate operations.
    
    The mapping allows the Er$^{3+}$ ions to be re-initialized so that new elementary links can be created between them. Fig.\ref{fig:main}(c) illustrates this processes for the $\mathrm{(i-1)}$th and $\mathrm{(i)}$th nodes. Immediately after generating entanglement between (the other) neighboring Er$^{3+}$ ions, the entanglement distance is extended by performing entanglement swapping between each of the closely-lying Er$^{3+}$ and $^{151}$Eu$^{3+}$ ions. As a result, the outer nodes becoming entangled. Fig. \ref{fig:main} (d) depicts a flow chart of our scheme. Photon loss and errors might cause some steps in the protocol to fail.
	
	As will be discussed in Sec. II A--C, the generation of entanglement between Er$^{3+}$ ions, the mapping of entanglement to the $^{151}$Eu$^{3+}$ ions, and the entanglement swapping steps, all rely on the detection of single photons that are spontaneously emitted from the Er$^{3+}$ ions. However, the lifetime of the $^4$I$_{15/2}$$\leftrightarrow$ $^4$I$_{13/2}$ telecommunication transition of Er$^{3+}$:Y$_2$SiO$_5$ is relatively long $T_1=11.4$ ms \cite{liu2006spectroscopic} with radiative lifetime being even longer $T_{rad}=54$ ms \cite{mcauslan2009strong}, which necessitates the need for a high-finesse cavity to enhance the emission rate. Nonetheless, the cavity will also significantly increase the probability of collecting the emitted photons as compared to emission into free space. We emphasize that the repetition rate of the protocol (i.e., the number of attempts of the protocol per unit time) is limited by the communication time between distant nodes, which means that there is no need for very fast emission. Next we discuss each of the steps of the protocol in detail.	
	\subsection{ Entanglement generation}\label{ssec:entanglement}
	Our scheme considers several remote cavities that each contain a single pair of close-lying Er$^{3+}$ and $^{151}$Eu$^{3+}$ ions. Note that there may be other RE ions within each cavity, but we assume that we can address a single such close-lying pair. See also section\ref{Sec:Implementation} below. An externally-applied magnetic field splits the degenerate electronic ground levels of Er$^{3+}$ via the Zeeman effect. We refer to the resultant $m_{s}=-\frac{1}{2}$ and $m_{s}=\frac{1}{2}$ Zeeman levels as the qubit states $\ket{\uparrow}$ and $\ket{\downarrow}$, respectively (see Fig. \ref{fig:main} (e)). 
	
	To generate entanglement between distant Er$^{3+}$ ions that are separated by a long distance $L_0$, we follow the scheme of Barret and Kok \cite{barrett,Hanson} (see also Fig. \ref{fig:main} (d)). First, each Er$^{3+}$ ion is prepared in one of the qubit states (here $m_s=-\frac{1}{2}$; denoted $\ket{\uparrow}$). After this initialization step, a $\frac{\pi}{2}$ microwave (MW) pulse rotates each Er$^{3+}$ ion into a superposition of $\ket{\uparrow}$ and $\ket{\downarrow}$ states. The application of a brief laser pulse that is resonant with the $\ket{\uparrow}$ $\leftrightarrow$ $\ket{e_2}$ transition, followed by spontaneous emission, will entangle each qubit state with the emitted photon number. That is, when the qubit state is $\ket{\uparrow}$ $(\ket{\downarrow})$ there will be $1$ $(0)$ emitted photon(s). The spontaneously-emitted photons are then directed to a beam splitter located in between the ions using optical fibers. A single-photon detection at one of the two beam splitter output ports projects the Er$^{3+}$ ions onto an entangled state.
	
	A possible loss in the fiber can lead to a situation where both Er$^{3+}$ ions emit a photon but only one photon is detected while the other is lost. In this case, the ions are left in a product state rather than an entangled state. To exclude this possibility, immediately after the first excitation-emission step of each Er$^{3+}$ ion, a $\pi$ MW pulse inverts each qubit state. Then a second excitation pulse is applied. The detection of two consecutive single photons at the beam splitter will leave the qubits in an entangled state:
	\begin{equation}
		\label{eq:Bellstate}
	\ket{\psi}_{\text{Er}}^{\pm}=\frac{1}{\sqrt{2}}(\ket{\uparrow\downarrow}\pm \ket{\downarrow\uparrow}).
	\end{equation}
    Here the + (-)  sign corresponds to the case in which the same (different) detector(s) received a photon. 
	\subsection{Controlled logic}
After successful entanglement of the two distant Er$^{3+}$ ions we transfer this entanglement to the neighboring $^{151}$Eu$^{3+}$ ions for a long-term storage. Efficient entanglement mapping between neighboring rare-earth ions can be employed by performing CNOT gate operations and single qubit rotations and read-out. Here we first explain the underlying mechanism and general scheme for how to implement a CNOT gate in our system, and in Sec.\ref{mappingsection} we will discuss in more detail how to swap entanglement between the ions using this gate.
	\begin{figure}
		\includegraphics[width=8.5cm]{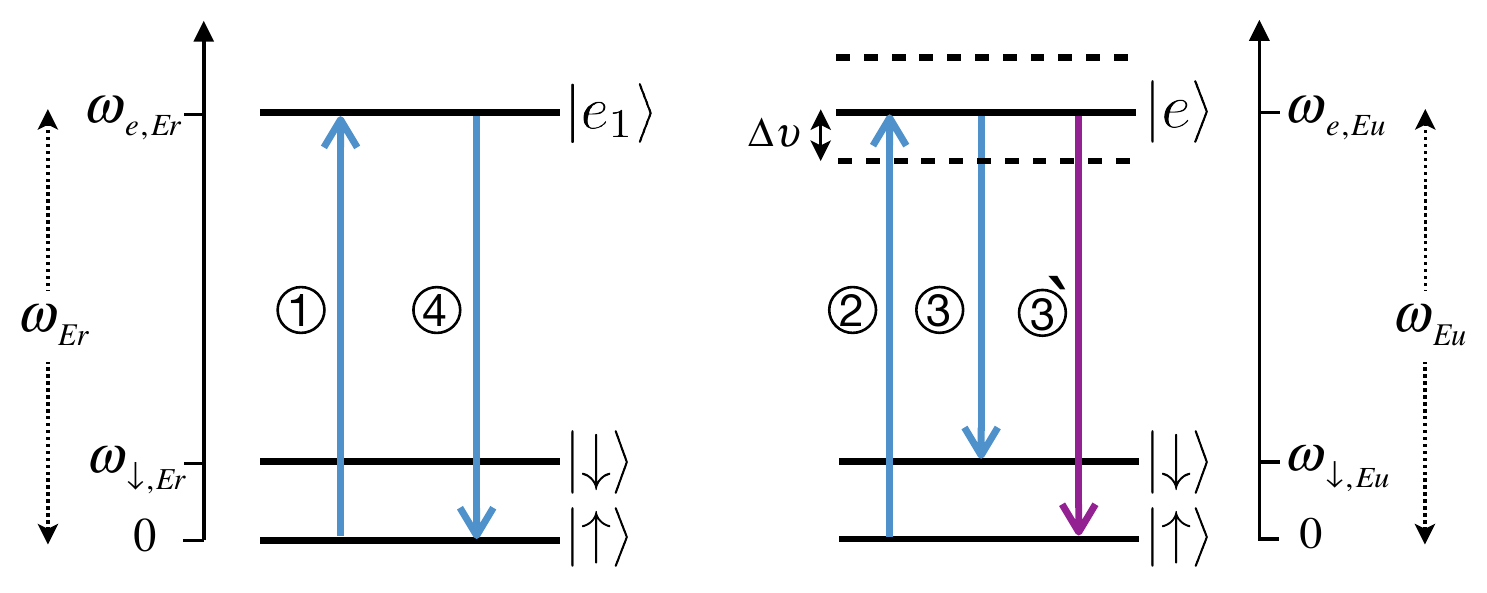}
		\caption{General pulse sequence of $\pi$-rotations to perform a controlled logic gate between nearby Er$^{3+}$ and $^{151}$Eu$^{3+}$ ions. The numbers indicate the sequential time ordering of the pulses. For this gate, the Er$^{3+}$ spin state acts as the control qubit and the $^{151}$Eu$^{3+}$ spin state acts as the target. A $\pi$ pulse excites the Er$^{3+}$ ion if it is in the state $\ket{\uparrow}$. When this occurs, pulses 2, 3 and 4 are not resonant with the $^{151}$Eu$^{3+}$ ion, leaving its state unaffected. Pulse 5 then brings the Er$^{3+}$ ion back to its original state. On the other hand, when Er$^{3+}$ is in the state $\ket{\downarrow}$, pulses 1 and 5 will be ineffective and hence it will remain in the ground state. Instead pulses 2, 3 and 4 will now be resonant with the $^{151}$Eu$^{3+}$ ion and (optically) swap its spin-state}
\label{fig:2}
	\end{figure}
Due to a lack of site symmetry when doped into a crystal, a RE ion can have a permanent electric dipole moment that is different depending on whether the ion is in its ground or optically excited state. The difference in the permanent dipole moments affects the optical transition frequency of other nearby RE ions via the Stark effect due to a modified local electric field environment. It is possible to dynamically control the shift in the transition frequency of one ion by optically exciting its neighboring ion. Based on this interaction, we can perform a controlled-NOT (CNOT) operation between nearby RE ions \cite{Kroll2002}. The modification of the transition frequency $\Delta\nu$ of a $^{151}$Eu$^{3+}$ ion by an Er$^{3+}$ ion due to the mutual electric dipole-dipole interaction is given by \cite{altner}:
	\begin{equation}
	\label{eq:shift}
\Delta\nu=\frac{\Delta\mu_\text{Er}\Delta\mu_\text{Eu}}{4\pi\epsilon\epsilon_{0}hr^{3}}\left(\left(\hat{\mu}_\text{Er}\!\cdot\!\hat{\mu}_\text{Eu}\right)-3\left(\hat{\mu}_\text{Er}\!\cdot\!\hat{r}\right)\left(\hat{\mu}_\text{Eu}\!\cdot\!\hat{r}\right)\right),
	\end{equation}
	where $r$ is the distance between the ions, and $\Delta\mu$ is the change of the permanent electric dipole moment of each ion, $h$ is the Planck constant, $\epsilon_{0}$ is vacuum permittivity, and $\epsilon$ is the dielectric constant.
	
 To perform a CNOT gate between the nearby Er (control) ion and Eu (target) ion, a sequence of five $\pi$ pulses is applied (see Fig.(\ref{fig:2})). First a $\pi$ pulse is applied to the Er$^{3+}$ ion on resonance with the $\ket{\uparrow}$ $\leftrightarrow$ $\ket{e_1}$ transition. From here, two cases must be considered: either the state of the Er$^{3+}$ ion is $\ket{\uparrow}$ or it is $\ket{\downarrow}$. 
 
If the Er$^{3+}$ ion is in state $\ket{\uparrow}$, it will be excited by pulse 1. This changes the permanent electric dipole moment of the Er$^{3+}$ ion, and thus its local electric field. Consequently, the transition frequency of the nearby $^{151}$Eu$^{3+}$ ion will be shifted by $\Delta\nu$. For the case that this frequency shift is large enough (that is, the $^{151}$Eu$^{3+}$ ion is sufficiently close to the Er$^{3+}$ ion), the $^{151}$Eu$^{3+}$ ion will be unaffected by pulses 2, 3 and 4, thereby remaining in its initial ground state (see Sec. \ref{cnotfidel} for more discussion). Finally, pulse 5 will bring Er$^{3+}$ ion back to its initial state. Hence, in this case, the pulse sequence does not modify the initial state of the ion pair system.
 
 On the other hand, if the Er$^{3+}$ ion is initially in $\ket{\downarrow}$, then pulses 1 and 5 will have no effect on the Er$^{3+}$ ion. Since the Er$^{3+}$ ion will not be excited, optical pulses 2, 3 and 4 will be resonant with the transitions of the $^{151}$Eu$^{3+}$ ion, and the pulse sequence will optically flip the two spin states of the $^{151}$Eu$^{3+}$ ion.
 \subsection{Entanglement mapping and distribution}\label{mappingsection}
  Both the ground and excited states of $^{151}$Eu$^{3+}$ have three doubly-degenerate nuclear hyperfine levels. With the application of a magnetic field, each doublet $\left(m_I=\pm\frac{1}{2},\pm\frac{3}{2},\pm\frac{5}{2}\right)$ will be split. For our proposal, we denote the $m_I=-\frac{3}{2}$ and $\frac{3}{2}$ hyperfine levels \cite{6hours} as the $^{151}$Eu$^{3+}$ ion qubit states, $\ket{\uparrow}$ and $\ket{\downarrow}$, respectively (see Fig. \ref{fig:main} (e)).
 	To map the state of each Er$^{3+}$ ion onto a nearby Eu ion, we first perform a CNOT gate between them. In our scheme, the $^{151}$Eu$^{3+}$ ion is initially prepared in one of the ground state levels (here $\ket{\uparrow}$) using optical pumping. In this special case, pulse 4 does not affect the system and can be neglected. This reduces the total gate time and improves the state mapping fidelity (see Sec. \ref{cnotfidel}).

After considering the phases that are acquired by performing the CNOT gate on the ions in neighboring nodes (labeled here as 1 and 2 instead of $\mathrm{(i)}$ and $\mathrm{(i+1)}$), the final state is
 		\begin{align}
 		\ket{\psi}_{\text{Er,Eu}}^{\pm} &= \frac{1}{\sqrt{2}} \left( \ket{\uparrow\downarrow}_{ \text{Er}_{1},\text{Er}_{2} }\ket{\uparrow\downarrow}_{\text{Eu}_{1},\text{Eu}_{2}} \nonumber  \right.\\ 
 		&\left.\pm e^{i\phi} \ket{\downarrow\uparrow}_{\text{Er}_{1},\text{Er}_{2}}\ket{\downarrow\uparrow}_{\text{Eu}_{1},\text{Eu}_{2} } \right),
 		\end{align} 
 		where 
 		\begin{align}
 		\phi &= (\omega_{\downarrow,\text{Er}_{2}}-\omega_{\downarrow,\text{Er}_{1}})\tau \nonumber + (\omega_{\text{Eu}_{2}}-\omega_{\text{Eu}_{1}})\tau_{2} \\ &+(\omega_{\downarrow,\text{Eu}_{2}}-\omega_{\downarrow,\text{Eu}_{1}}) (\tau_{3}+\tau_{4}) \nonumber \\ &+ k_{\downarrow \text{Eu}_{1}}x_{\text{Eu}_{1}}-k_{\downarrow \text{Eu}_{2}}x_{\text{Eu}_{2}},
 		\end{align} 
 		and $x$ is the distance the photon travels between each $^{151}$Eu$^{3+}$ ion and the beam splitter, $k_{\downarrow \text{Eu}}=\frac{\omega_{\downarrow,\text{Eu}}}{c}$ is the wavenumber, $\tau_{j}$ is the time elapsed after application of the $j^{\text{th}}$ pulse and $\tau$ is the total time duration that is needed to perform a CNOT gate. Since these phases are known, they can be compensated by applying local operations on the ions.
 		
To conclude the mapping step, a $\frac{\pi}{2}$ microwave (MW) pulse is applied to rotate each Er$^{3+}$ qubit. This is followed by a state measurement of both nearest-neighbor Er$^{3+}$ ions, see also Sec. IV below. This projects the $^{151}$Eu$^{3+}$ ions onto an entangled state. Depending on the outcome of these measurements, the entangled state between remote $^{151}$Eu$^{3+}$ ions would be $\ket{\psi^+}$ or $\ket{\psi^-}$. For example, in the case that we begin with $\ket{\psi}_{\text{Er}}^{+}$ (given in Eq.~\ref{eq:Bellstate}), after performing the gate, MW rotation, and measurement, if both Er ions are found in the state $\ket{\uparrow}$, the entangled state between remote $^{151}$Eu$^{3+}$ ions is $\frac{1}{\sqrt{2}}\left(\ket{\uparrow\downarrow}_{\text{Eu}_{1},\text{Eu}_{2}}+ \ket{\downarrow\uparrow}_{\text{Eu}_{1},\text{Eu}_{2}}\right)$.
 	
	Once entanglement is established in neighboring elementary links, we perform a joint measurement on both ions at each intermediate node to distribute entanglement (see Fig. \ref{fig:main} (c)). In our scheme, it is possible to perform entanglement swapping deterministically using the permanent electric dipole-dipole interaction.
	
	To perform the desired entanglement swapping, first a CNOT gate is applied in which Er$^{3+}$ serves as the control qubit and $^{151}$Eu$^{3+}$ as the target qubit (similarly as before). Here, both ions are in a superposition state and so all 5 pulses of the CNOT gate sequence are required. Then the Er$^{3+}$ ion is measured in the diagonal (X) basis. Next, another CNOT operation is performed but now the target and control qubits are exchanged. Finally the Er$^{3+}$ ion is measured in the logical (Z) basis. This `reverse' CNOT gate is performed in order to avoid directly measuring the spin state of $^{151}$Eu$^{3+}$ ion optically (the optical lifetime of $^{151}$Eu$^{3+}$ is longer than the Er$^{3+}$ spin coherence lifetime). Fig.(\ref{fig:2}) shows the pulse sequence needed to perform the first CNOT gate between the ions.
    
	Based on the outcomes of the measurements on each Er$^{3+}$ ion, the outer nodes will be projected onto one of the four Bell states. To be more precise, when performing entanglement swapping between $\mathrm{Er_{i}}$ and $\mathrm{Eu_{i}}$, if only one of the state measurements of $\mathrm{Er_{i}}$ is $\ket{\uparrow}$, the state of $\mathrm{Er_{i-1}}$ and $\mathrm{Eu_{i+1}}$ will project onto the $\ket{\psi^+}$ or the $\ket{\psi^-}$ Bell state. On the other hand, if both measurements of $\mathrm{Er_{i}}$ are $\ket{\uparrow}$ or both $\ket{\downarrow}$, the state will project onto the $\ket{\phi^+}$ or the $\ket{\phi^-}$ Bell state.

 To verify or utilize the entanglement that is generated between the distant ions, a measurement must be performed on the endpoint nodes. Our protocol results in an entangled state of the Er$^{3+}$ ion at the first node with the $^{151}$Eu$^{3+}$ ion at the last node. A measurement of the Er$^{3+}$ ion can be performed by direct optical excitation. A measurement of the $^{151}$Eu$^{3+}$ ion spin can be performed by using the nearby Er$^{3+}$ as a readout ion. This is done by mapping the $^{151}$Eu$^{3+}$ state to Er$^{3+}$ using gate operations (also see section \ref{statemeasurement} for further details).
 \color{black}
    \section{Entanglement generation rates and multiplexing}\label{Sec:rate}
	
	In our scheme, the success probability of generating an entangled state between two neighboring Er$^{3+}$ ions that are separated by a distance $L_{0}$ is $p_{t}=\frac{1}{2} \eta_{t}^2 p^2 \eta_{d}^2$, where $\eta_{t}=e^{-\frac{L_{0}}{2L_{att}}}$ is the transmission probability of a photon through optical fiber and $L_{att}\approx22$ km (which corresponds to a loss of 0.2 dB$/$km), $\eta_{d}$ is the detection efficiency, and $p$ is the success probability of emitting a single photon into a cavity mode. Similarly, the success probability of mapping the state of an Er$^{3+}$ ion onto a nearby $^{151}$Eu$^{3+}$ ion is $p_m\approx p \eta_{d}$, because it requires one optical read-out of each Er$^{3+}$ ion. Therefore, the average time to generate entanglement in an elementary link and perform the state mapping steps is 
	\begin{equation}
		\langle T\rangle_{L_0}= \frac{L_0}{c} \left(\frac{1}{p_{t}p_m^2}\right),
	\end{equation}
	where $c=2\times10^8 \mathrm{\frac{m}{s}}$ is the speed of light in fiber.
	
	The average time that is required to distribute entanglement over two neighboring elementary links, which corresponds to a distance $L=2L_{0}$, can be estimated as follows. First, entangled pairs of $^{151}$Eu$^{3+}$ ions are generated over an elementary link and, once successful, Er$^{3+}$ ions are then generated in a neighboring elementary link. Thus, the probability of establishing entanglement for both links is
	\begin{equation}
		p_0= \left(\frac{1}{P_A}+\frac{1}{P_B}\right)^{-1},
	\end{equation}
		where $P_A=p_{t}$ and $P_B=p_{t}p_m^2$ represent the success probability of generating entanglement between Er$^{3+}$ and $^{151}$Eu$^{3+}$ ions, respectively.	
	 Then, entanglement is extended by performing a BSM between the close-lying Er$^{3+}$ and $^{151}$Eu$^{3+}$ ions at the center node.
	The success probability for the entanglement swapping step is $p_s\approx  p^2 \eta_{d}^2$ because it requires two optically-induced spin read-outs of an Er$^{3+}$ ion. Consequently, the average time to distribute entanglement over a distance $2L_0$ is
	\begin{equation}
		\braket{T}_{2L_0}=\frac{L_0}{c}\frac{1}{p_0p_s}.
	\end{equation}
	
	Accordingly, the average time to distribute an entangled pair over a distance $L=2^nL_0$, with $n$ denoting the number of nesting levels, is \cite{singletrappedions,review-multi1}:
	\begin{equation}
		\braket{T}_{L}=\left(\frac{3}{2}\right)^{n-1}\frac{L_0}{c} \frac{1}{p_0  p_s^n}.
	\end{equation}
	The factor $3/2$ for each of the next nesting levels, which is a good approximation for the exact result \cite{review-multi1}, can be understood in following way. 
    
    In contrast to the first nesting level, a successful entanglement distribution for higher nesting levels does not require waiting for a success in one link before the establishment of entanglement in another link can be attempted; rather, the establishment of entanglement can be attempted in both links simultaneously. For example, if a goal is to distribute entanglement over the distance $4L_0$, entanglement must be generated in two neighboring links each of length $2L_0$ before entanglement swapping is performed. If the average waiting time for a success in one link of length $2L_0$ is $\braket{T}_{2L_0}$, entanglement will be established in one of two links after $\braket{T}_{2L_0}/2$. Then, after another $\braket{T}_{2L_0}$ time duration, entanglement will be established in the other link. Hence, the average time to establish entanglement in the two neighboring links is $3\braket{T}_{2L_0}/2$. The same arguments hold for the next nesting levels, resulting in a factor $\left(\frac{3}{2}\right)^{n-1}$ for $n$ nesting levels.
    
\begin{figure}
		\includegraphics[width=8.5cm]{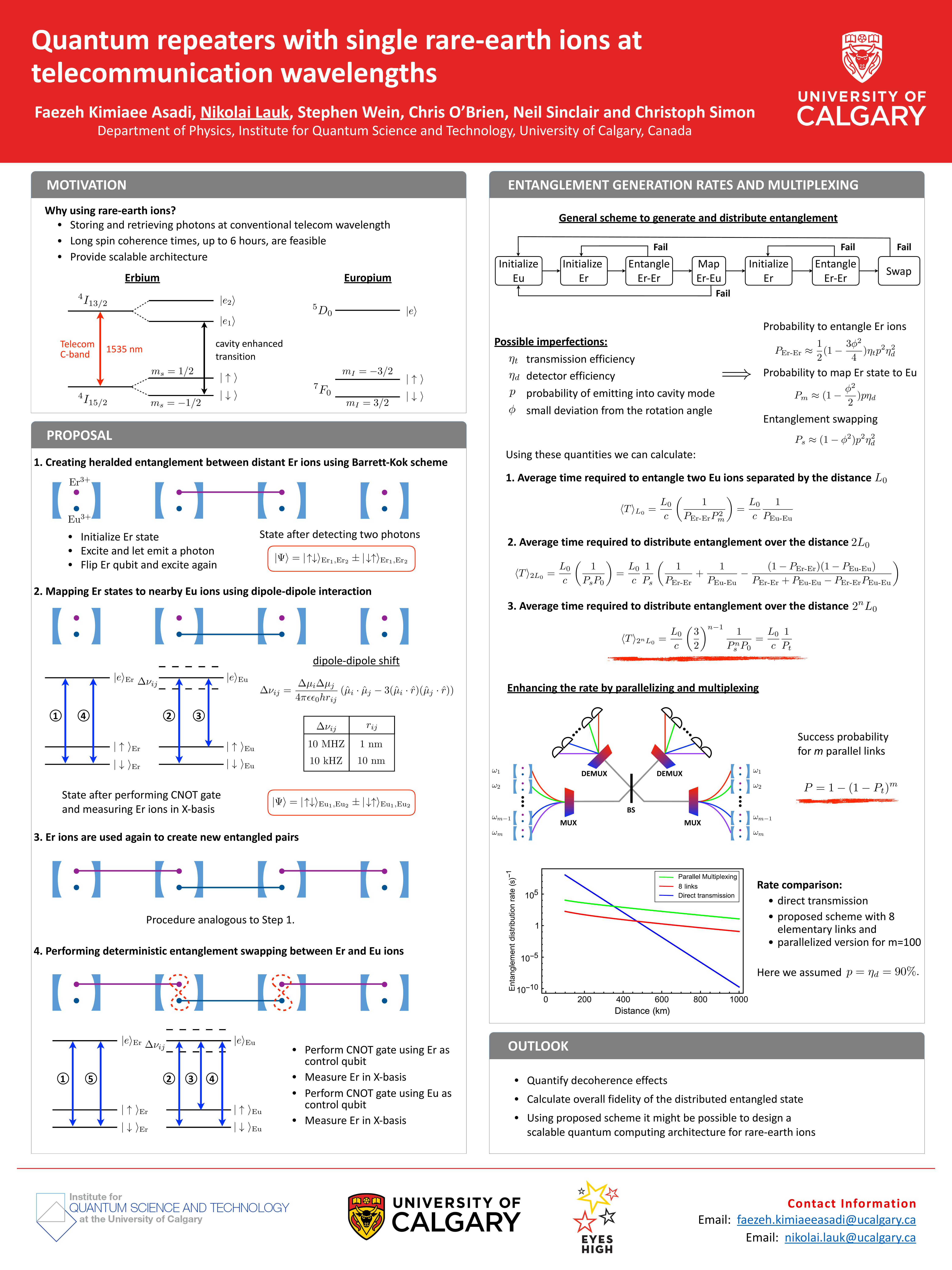}
		\caption{Our multiplexed scheme consists of $2^n$+1 nodes that span the total channel distance $L$. Each node consists of an array of $m$ cavities that emit photons which feature differing carrier frequencies. The carrier frequency of each photon is determined by a frequency translation device. A coupling element (COUPLER) ensures that each photon traverses a common channel to a Bell-state measurement station that consists of a beam splitter, decoupling element (DECOUPLER), and 2$m$ single photon detectors. This set-up allows $m$ entanglement generation protocols to be operated in parallel.
		}\label{fig:multiple}
	\end{figure}

	Multiplexing can be employed to significantly enhance the rate of entanglement distribution. Referring to the encoding of several individual qubits, each into their own distinguishable modes, multiplexing has been utilized for some quantum repeater proposals \cite{review-multi1,entanglementgenerationtime1, averagetime,Neil}. As outlined in Fig. \ref{fig:multiple}, we consider an array of $m$ cavities in which each cavity emits a photon that features a distinguishable resonance frequency from the rest. This can be accomplished using frequency translation (see Sec. IV for more details). Using a coupler device, each photon is directed into a common fiber and is directed towards a BSM station that is composed of similarly-designed decouplers and single-photon detectors. Benefiting from the distribution of entanglement into $m$ parallel modes, the probability that at least one entangled state is distributed over the entire channel is $1-(1-P_t)^m$, which can be made unity for a sufficiently high $m$. Here, $P_t=(2/3)^{n-1} p_0 p_s^n$ is the success probability of distributing entanglement over a distance $L$ using a qubit that is encoded into a single mode. Further details of this set-up are described in Sec.\ref{ssec:multiplexed}.
	
\begin{figure}	\includegraphics[width=0.5\textwidth]{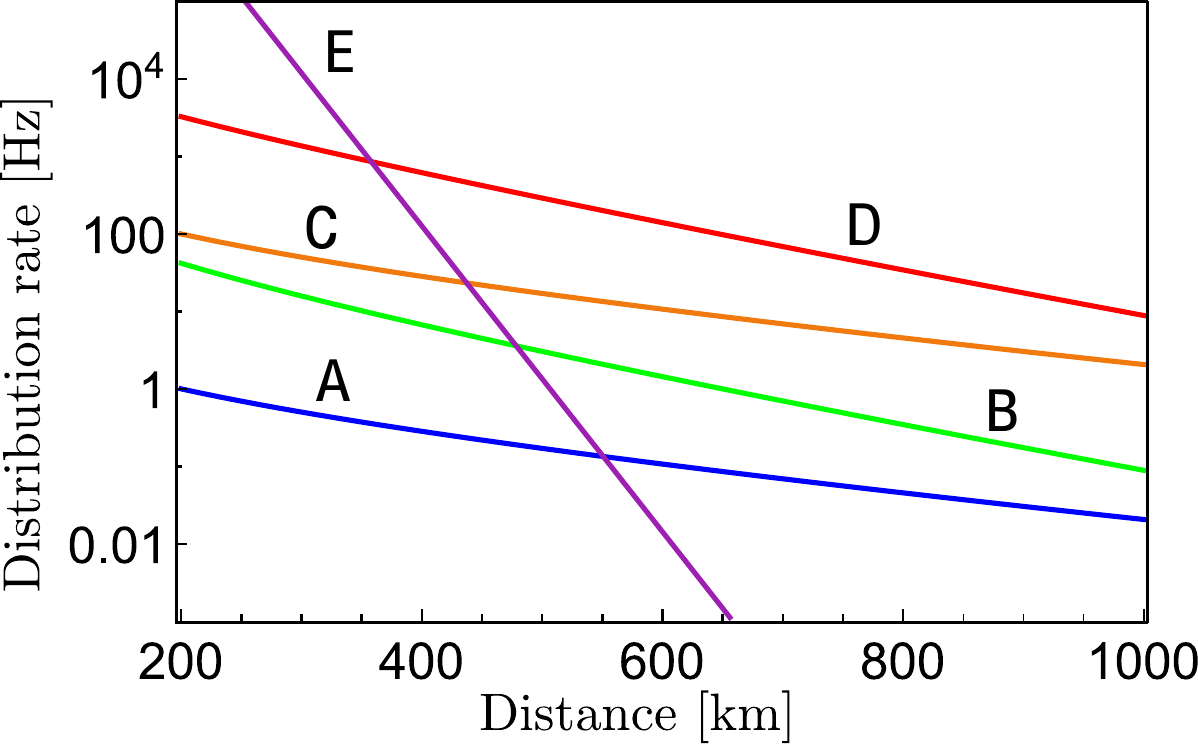}
		\caption{A comparison of the entanglement distribution rate for various schemes. Direct transmission scheme using a 10 GHz single-photon source (E) is compared with original DLCZ protocol (A) and our scheme (B) using 8 elementary links, each of length $L_0$. Also shown are the rates that correspond to the multiplexed versions of DLCZ (C) as well as our scheme (D) with $m=100$ using the same number of elementary links. We assume that the detection efficiency and the success probability of emitting a single photon by an ion is $p=\eta_{d}=0.9$, and in the case of DLCZ, a storage efficiency of $\eta_m =0.9$. The repetition rate for each repeater protocol is set by the communication time $L_0/c$.}\label{fig:rate}
	\end{figure}

	The entanglement distribution rate of our scheme is plotted as a function of distance for $n=3$ in Fig. \ref{fig:rate} and compared to that employing the well-known DLCZ scheme \cite{DLCZ}, which uses ensemble-based memories, as well as that using direct transmission with a single-photon source which produces photons at 10 GHz. This 10 GHz photon rate is an optimistic rate for the direct transmission of photons. 
The rate that we assume for a single photon source is much faster than the rate for the proposed repeater because they have different limitations. Even though the photon rate of our scheme (which depends on the cavity characteristics and the optical lifetime of the Er$^{3+}$ ion) is much lower, the time scale for the repeater is actually determined by the communication time L$_0/c$ which is even longer. Note that the direct transmission scheme can also be interpreted as the Pirandola bound \cite{Pirandola} for a repetition rate of 10/1.44$=$6.9 GHz. 
    The performance of our protocol and the DLCZ scheme with $m=100$ multiplexed channels is also shown Fig. \ref{fig:rate}. For more information on other approaches see Ref. \cite{review-multi1}. In this review paper, the multiplexed DLCZ outperformed the other repeater protocols considered (see Fig. 18 of ref.\cite{review-multi1}). As will be discussed in Sec. IV B, the use of even more than $m=100$ parallel spectral channels is possible.
    
  Fig. \ref{fig:rate} shows that our use of deterministic gates is advantageous for the rate. The scaling of the DLCZ scheme with distance is slightly better than that of our scheme due to the requirement of detecting only one photon for each elementary entanglement creation step (rather than two photons for our scheme). However, this comes at the significant expense of requiring phase stabilization for the long-distance fiber links.
\section{Implementation}\label{Sec:Implementation}
	For our scheme, we consider ion beam-milled Y$_2$SiO$_5$ photonic crystal cavity systems that have been weakly-doped with Er$^{3+}$ and $^{151}$Eu$^{3+}$ ions \cite{faraon,faraon2,cone}. Co-doped crystals may be grown from the melt \cite{cone}, or individual RE ions may be implanted into single Y$_2$SiO$_5$ crystals \cite{implant}. After a milling step, the output of the cavity can be coupled to an optical waveguide using, e.g., microscopy \cite{davanco}, bonding \cite{murray} or a pick-and-place technique \cite{kim}, with the latter having been used to heterogeneously interface InAs/InP quantum dots with Si waveguides. Despite the lack of on-demand control of the position and relative orientation of each ion \cite{implant}, a suitable Er-Eu ion pair can be identified by performing laser-induced fluorescence spectroscopy. This involves exciting each Er$^{3+}$ ($^{151}$Eu$^{3+}$) ion with a laser and measuring the spectral shift of the resultant fluorescence from other $^{151}$Eu$^{3+}$ (Er$^{3+}$) ions due to the aforementioned electric-dipole coupling (see Sec. \ref{mappingsection}) \cite{cone}. Note that measurements using a 0.02\%Er:1\%Eu:Y$_2$SiO$_5$ bulk crystal revealed the optical transition frequency of sets of Er$^{3+}$ ions that lie within approximately one nanometer from adjacent Eu$^{3+}$ ions \cite{cone}. These results suggest that the transition frequencies of suitable Er-Eu pairs can be rapidly distinguished from other spectator ions.
    
  The magnetic field applied in the $D_1-D_2$ plane at 135 degrees relative to the $D_1$ axis results in the decay of the excited Er$^{3+}$ ion back into the initial spin state via spontaneous emission with a probability of higher than 90\%\cite{Zeeman}, and a Zeeman-level lifetime of about 130ms was measured for a magnetic field of 1.2 mT at 2.1 K temperature\cite{Zeeman}. At large external magnetic fields of 1 T or more and temperatures below 3 K, one-phonon direct process is the dominant spin-relaxation mechanism and the temperature and magnetic field dependence of the relaxation rate could be approximated by \cite{bottger}:
\begin{equation}   R(B)=R_{0}+\alpha_{D}g^3B^5\text{coth}(\frac{g\mu_{B}B}{2kT})
    \end{equation}
where $\alpha_D$ is the anisotropic constant, $\mu_{B}$ is the Bohr magneton, and $k$ is the Boltzmann constant. The field independent contribution $R_{0}$ can be attributed to cross-relaxations with paramagnetic impurities in the crystal and hence depends strongly on the crystal purity.  
Extrapolating from Refs \cite{bottger,Zeeman}, the spin relaxation time would be about 40ms at 20mK for an external magnetic field of 1T at the 135$^{\circ}$ in $D_1-D_2$ plane. At this magnetic field, however, the splitting of the Zeeman levels would be too large to address it with microwave pulses and optical Raman pulses should be applied instead.

For an ensemble-doped Er$^{3+}$:Y$_2$SiO$_5$ the spin coherence lifetime at mK temperatures and external magnetic fields of few hundred mT can be as short as $\sim 7\hspace{1mm}\mu$s \cite{Bushev}. However, for an ensemble-doped Er$^{3+}$:Y$_2$SiO$_5$ crystal spin flip-flop processes are one of the main sources of decoherence, the spin coherence lifetime of a single Er$^{3+}$ ion in Y$_2$SiO$_5$ is largely determined by spin-spin magnetic dipole interactions. The magnetic moments of the constituent spins of Y$_2$SiO$_5$ are small: $-0.137\mu_{N}$, $-0.5\mu_{N}$, and $-1.89\mu_{N}$ for $\mathrm{^{89}Y}$, $\mathrm{^{29}Si}$, and $\mathrm{^{17}O}$ respectively. Compared to yttrium ions, $\mathrm{^{29}Si}$ and $\mathrm{^{17}O}$ have low isotopic natural abundances, so we assume the contribution of these isotopes to be negligible and only consider the Er-Y interactions. In a large enough magnetic field (a few hundred milli-tesla or more) compared to the Er-Y coupling strength, the magnetic moment of Er$^{3+}$ will detune the closest-proximity Y ions from resonance with those further away. This well-known effect referred to as the  ``frozen core'' has been observed and results in weaker decoherence of the Er$^{3+}$ ion by nearby Y ions \cite{Direct,pr-y,frozencore}.
   
 Hence, for a single Er$^{3+}$ in Y$_2$SiO$_5$, the presence of a strong magnetic field may increase the spin coherence lifetime into the milliseconds range. To further increase the spin coherence lifetime, dynamical decoupling immediately after the second optical excitation step is necessary. While this has not been demonstrated in Er$^{3+}$:Y$_2$SiO$_5$, it is a widely-employed method, and has been used to extend coherence lifetimes in Ce-doped yttrium aluminum garnet \cite{Ce} as well as nitrogen-vacancy centers \cite{NV-Carbon}. 
    
    Spin polarization of up to $90\%$ in Er$^{3+}$:Y$_2$SiO$_5$ has been realized by using stimulated emission and spin-mixing methods \cite{statepreparation}. The efficiency of spin polarization is determined by a competition between decay of the ground-level population relative to the optical pumping efficiency of the ion.  Since the latter will be enhanced due to the cavity-induced decay rate, we expect a near-unity Er$^{3+}$-spin polarization for our scheme. Since the ground state lifetime of $^{151}$Eu$^{3+}$ is several hours, spin polarization can be achieved by performing continuous optical pumping of all but one of the ground states. 
	
    In Y$_2$SiO$_5$, RE ions can occupy two crystallographically inequivalent sites with $C_{1}$ symmetry. Due to this lack of symmetry, the orientation of the dipole moments and their magnitudes, and hence their dipole-dipole interaction strengths, are unknown. 
    Previous measurements of the linear Stark shift of Er$^{3+}$:Y$_2$SiO$_5$ \cite{CRIB} allow the calculation of the projection of the electric-dipole moment difference onto the direction of the externally applied electric field to be approximately $0.84\times10^{-31}$Cm \cite{macfarlane}. The dipole moment difference for $^{151}$Eu$^{3+}$:Y$_2$SiO$_5$ can be as high as $\Delta\mu_{Eu}=0.81\times10^{-31}$Cm \cite{EU-dipolemoment}, resulting in the shift of the transition frequency of $10$ and $0.01$ MHz for $r_{ij}=1$ and $10$ nm, respectively. As a comparison, the magnetic dipole-dipole interaction between Er$^{3+}$ magnetic moment, which can be as high as $\mu_{Er}=14.65\ \mu_B$ \cite{Sun2008}, and $^{151}$Eu magnetic moment with its intrinsic value of $\mu_{Eu}=3.42\ \mu_N$ \cite{Eu-magnetic} is much weaker and amounts to 363 and 0.363 kHz for $r_{ij}=1$ and $10$ nm, respectively.
	\subsection{Cavity}
	\label{cavitysection}
	The cavity serves three main purposes in this proposal. It improves the quantum efficiency, enhances the single-photon indistinguishability, and increases the rate of Er$^{3+}$ emission. To achieve these benefits, the cavity must provide a significant Purcell enhancement to the $\ket{e_2}\longrightarrow\ket{\uparrow}$ transition of Er$^{3+}$.

  In the context of RE ions, Purcell factors of several hundred have been achieved  \cite{Thompson,faraon}, and up to $10^3$ seems to be a reasonable goal \cite{Obrien}. The Purcell factor can be written: $P = (\gamma/\gamma_r)C$, where $1/\gamma$ is the excited state lifetime, $\gamma_r$ is the radiative decay rate, and $C$ is the cavity cooperativity. The cavity-enhanced quantum efficiency (probability of emitting a photon into the cavity mode) is then given by: $p= \eta P/(1+\eta P)$, where $\eta=\beta\gamma_r/\gamma$ is the Er$^{3+}$ spin-conserving quantum efficiency, and $\beta$ is the probability of an excited ion to relax into the initial spin state via the spontaneous emission. For Er$^{3+}$:Y$_2$SiO$_5$,  $\beta=0.9$, $\gamma=2\pi\times 14$ Hz, and $\gamma_r=2\pi\times 3$ Hz \cite{mcauslan2009strong}, resulting in $\eta=0.19$. For $P=100$, a cavity quantum efficiency of $p=0.95$ is possible. Increasing the Purcell factor to $P=1000$ allows $p=0.995$.
	
The single-photon indistinguishability is a metric that quantifies the quality of interference between photons that originate from the same quantum emitter. Without a cavity, the single-photon indistinguishabilty can be defined as $I_1=T_2/2T_1$ \cite{grange2015}. For an ensemble of 0.0015\% Er$^{3+}$:Y$_2$SiO$_5$ we have $T_1=11.4$ms \cite{liu2006spectroscopic} and optical $T_2$ is around 200$\mu$s at 1 T and a temperature of a few K \cite{4ms}.  This would imply $I_1=0.009$ without a cavity, which would require significant spectral filtering to achieve successful entanglement generation. With a cavity, the single-photon indistinguishability can be approximated by: $I_1=(1+\eta P)/(\zeta+1+\eta P)$, where $\zeta=2T_1/T_2^\star=2T_1/T_2-1$ is the dephasing ratio. Therefore, with Purcell factors $P=1000$ and $P=20,000$ the single-photon indistinguishability would be $I_1=0.63$ and $I_1=0.97$, respectively. In a large magnetic field, however, $T_2=4$ ms was measured at a few K \cite{4ms}. For an ensemble-doped Er$^{3+}$:Y$_2$SiO$_5$ spin flip-flop processes are the dominant decoherence mechanism and applying a large magnetic field can freeze these processes. As a result we have a long optical $T_2$. The flip-flop process can also be suppressed by reducing the Er$^{3+}$ concentration. Therefore in our system, where we assume very low dopant concetration, we still can expect to have optical coherence times of the order of a few ms even at lower magnetic fields (hundreds of mT). With $P=100$ and $T_2=4$ms, $I_1=0.82$ is possible, which could be further improved by attempting to spectrally filter the narrow $2\pi\times 1.4$ kHz bandwidth photons. With $P=1000$, $I_1=0.98$ could be achieved without spectrally filtering the $2\pi\times 14$ kHz photons (corresponding to a duration of 11 $\mu$s).
	
	\subsection{CNOT gate}
    \label{cnotfidel}
	
	To perform the CNOT gate, it is necessary for Er$^{3+}$ to remain excited for a time that is long enough to apply three $\pi$-pulses to $^{151}$Eu$^{3+}$. This implies that each Er$^{3+}$ ion must be excited to a different Zeeman level ($\ket{e_1}$) than the level that is coupled to the cavity ($\ket{e_2}$). This can be done if the Zeeman splitting between $\ket{e_1}$ and $\ket{e_2}$ is much larger than the cavity linewidth. In this section, for Er$^{3+}$  we use $\ket{e}_\text{Er}=\ket{e_1}$.

	To quantify the fidelity of the CNOT gate and state transfer procedures, we use a model nine-level system where both Er$^{3+}$  and $^{151}$Eu$^{3+}$  are represented by three-level systems with two lower-energy spin states ($\ket{\uparrow}$, $\ket{\downarrow}$) and a single excited state ($\ket{e}$) (see Fig. \ref{fig:2}). We use the dipole-interaction Hamiltonian:
	\begin{equation}
		H = \Delta\nu\sigma^+_{\text{Er}\uparrow}\sigma^-_{\text{Er}\uparrow}\sigma^+_{\text{Eu}\uparrow}\sigma^-_{\text{Eu}\uparrow}+\sum_{k,l}\frac{\Omega_{k,l}}{2}\left(\sigma^+_{k,l}+\sigma^-_{k,l}\right),
	\end{equation}
	where $\Delta\nu$ is the dipole interaction strength and $\Omega_{k,l}$ is the Rabi frequency for transition $\{k,l\}$ for $l\in\{\uparrow,\downarrow\}$, $k\in\{\text{Er},\text{Eu}\}$. Without loss of generality, we choose $\Delta\nu>0$ and $\Omega_{k,l}>0$. The master equation is
	\begin{equation}
		\begin{aligned}
			\dot{\rho} &= -i[H,\rho]+\sum_{k,l}\gamma_{k,l}\mathcal{D}(\sigma_{k,l}^-)\rho+\frac{\gamma^\star_k}{2}\mathcal{D}(\sigma^+_{k,l}\sigma^-_{k,l})\rho\\
			&+\sum_k\frac{\chi_k}{2}\mathcal{D}(\sigma_{z,k})\rho,
		\end{aligned}
	\end{equation}
	where $\mathcal{D}(\sigma)\rho=\sigma\rho\sigma^\dagger-\sigma^\dagger\sigma\rho/2-\rho\sigma^\dagger\sigma/2$. The operators are defined as $\sigma_{k,l}^+=\ket{e}\bra{l}_k$, $\sigma_{k,l}^-=(\sigma_{k,l}^+)^\dagger$, and $\sigma_{z,k} = (\ket{\uparrow}\bra{\uparrow}_k-\ket{\downarrow}\bra{\downarrow}_k)/2$. The rate $\gamma_{k,l}$ is the decay rate for transition $\{k, l\}$, $\gamma^\star_k$ is the optical pure dephasing rate, and $\chi_k$ is the spin decoherence rate for ion $k$.
	
	To solve for the Er-Eu state after applying the five-pulse CNOT gate sequence, we first assume that each $\pi$-pulse is a square pulse and that they are applied to the system sequentially with no time delay between pulses. In this case, the time taken to apply each $\pi$-pulse is given by $T_{k,l}=\pi/\Omega_{k,l}$. We also assume that $\Omega_{k,l}$ and $\Delta\nu$ are much larger than any dissipative rate so that we can treat the dissipation perturbatively.
    
    We then define the zero-order (reversible) superoperator $\mathcal{L}_{0}$ where $\mathcal{L}_{0}\rho = -i[H,\rho]$. Then we define the first-order (irreversible) perturbation superoperator $\mathcal{L}_{1}$ as $\mathcal{L}_{1}\rho=\dot{\rho}-\mathcal{L}_{0}\rho$. From this, we define the rotation superoperator $\mathcal{R}_{k,l}(\theta)$ corresponding to the pulse where $\Omega_{k,l}\neq 0$ and $\Omega_{k^\prime,l^\prime}=0$ for all $k^\prime\neq k$ and $l^\prime\neq l$. This superoperator is given by
	\begin{equation}
		\mathcal{R}_{k,l}(\theta) = e^{\mathcal{L}_{0} \frac{\theta}{\Omega_{k,l}}}\left[1 + \int_0^{\frac{\theta}{\Omega_{k,l}}}d\tau e^{-\mathcal{L}_{0}\tau}\mathcal{L}_{1}e^{\mathcal{L}_{0} \tau}\right],
	\end{equation}
	which is accurate to first-order in $\gamma_{k^\prime,l^\prime},\gamma^\star_{k^\prime},\chi_{k^\prime}\ll \Omega_{k,l},\Delta\nu$ for all $k^\prime$ and $l^\prime$.
	
	To simulate the entanglement swapping between $^{151}$Eu$^{3+}$  and Er$^{3+}$, we begin with an initial superposition state $\rho_0$ where $\ket{\psi}=(\ket{\uparrow}+\ket{\downarrow})_\text{Er}(\ket{\uparrow}+\ket{\downarrow})_\text{Eu}$ and $\rho_0=\ket{\psi}\bra{\psi}$. Then after applying the CNOT pulse scheme, the final state is
	\[
	\rho=\mathcal{R}_{\text{Er},\uparrow}(\theta)
	\mathcal{R}_{\text{Eu},\uparrow}(\theta)
	\mathcal{R}_{\text{Eu},\downarrow}(\theta)
	\mathcal{R}_{\text{Eu},\uparrow}(\theta)
	\mathcal{R}_{\text{Er},\uparrow}(\theta)\rho_0,
	\]
	where ideally $\theta=\pi$; however, we set $\theta=\pi+\epsilon$ to simulate a $\pi$-pulse with a small rotation deviation $|\epsilon|\ll\pi$ caused by a non-ideal pulse area. The total time taken to apply this sequence is $T_\text{gate}=2T_{\text{Er},\uparrow}+2T_{\text{Eu},\uparrow}+T_{\text{Eu},\downarrow}$.
	
	For a high fidelity gate, the pulse durations must be small so that the gate is fast and qubit states do not dephase during the sequence. This favors larger Rabi frequencies. On the other hand, it is also necessary that the three $\text{Eu}$ pulses do not excite the $^{151}$Eu$^{3+}$  spin state if Er$^{3+}$ is excited. One way to achieve this is to set $\Omega_{\text{Eu},l}$ to satisfy the detuning condition $\Omega_{\text{Eu},l}^2/\Delta\nu^2\ll 1$ so that there is little chance for off-resonant excitation. However, since $\Delta\nu$ cannot be arbitrarily large (the separation between Er and Eu cannot be arbitrarily small), the detuning criteria necessitates a very slow gate, which cannot provide a high fidelity.
    
Alternatively, the detuning condition can be circumvented if $\Omega_{\text{Eu},l}$ is chosen so that the Eu pulses perform an effective $2\pi$ rotation on the Eu spin state when Er is excited, but still perform a $\pi$-pulse when Er is not excited. This can be accomplished by requiring that $\Omega_{\text{Eu},l}$ satisfies the effective Rabi frequency relation $\sqrt{\Delta\nu^2+\Omega_{\text{Eu},l}^2}=2\Omega_{\text{Eu},l}$. This sets $\Omega_{\text{Eu},l}=\Delta\nu/\sqrt{3}$. As a consequence of fixing the Rabi frequency, an accurate characterization of the dipole interaction strength for each Er-Eu pair is necessary to achieve a high fidelity. This is because any mischaracterization $\delta\nu$ from the true value $\Delta\nu-\delta\nu$ will cause a deviation from the desired $2\pi$ rotation. To account for this, we also consider a perturbation of the fidelity for deviation $\Delta\nu\rightarrow\Delta\nu-\delta\nu$ where we assume $|\delta\nu|\ll\Delta\nu\propto\Omega_{k,l}$.
    
	The effective $2\pi$ pulse leaves a relative phase between the $^{151}$Eu$^{3+}$ ground-state spins of $\varphi = -\pi(2-\sqrt{3})/2$. If Er$^{3+}$ is in $\ket{\downarrow}$, then only $\pi$-rotations are applied to $^{151}$Eu$^{3+}$. Hence, in this case, there is no acquired relative phase from the detuned pulses. However, if the state is $\ket{\uparrow\downarrow}$, $^{151}$Eu$^{3+}$ will be affected by the third pulse performing an effective $2\pi$ rotation, and so $\ket{\uparrow\downarrow}$ will acquire a relative phase of $\varphi$. Likewise, if the state is $\ket{\uparrow\uparrow}$, $^{151}$Eu$^{3+}$ will be affected by the second and fourth pulses performing an effective $2\pi$ rotation, and so $\ket{\uparrow\uparrow}$ will acquire a relative phase of $2\varphi$. Hence, in the absence of dissipation and using perfect square $\pi$-pulses, the expected final state is
	\begin{equation}
\ket{\psi_f}=\frac{1}{2}\left(\ket{\downarrow\downarrow}+\ket{\downarrow\uparrow}+e^{i\varphi}\ket{\uparrow\downarrow}+e^{i2\varphi}\ket{\uparrow\uparrow}\right)
	\end{equation}
    Since the expected acquired phase $\varphi= -\pi(2-\sqrt{3})/2$ is known, and independent of the dipole interaction strength $\Delta\nu$, it can be tracked or corrected and so we use $\ket{\psi_f}$ as the final state when calculating the fidelity.
    
	We use the above expressions for $\rho$ and $\ket{\psi_f}$ to compute the fidelity $F_\text{CNOT}=|\bra{\psi_f}\rho\ket{\psi_f}|$. For simplicity, we choose to set $\Omega_{\text{Er},l}=\Omega_{\text{Eu},l}=\Omega=\Delta\nu/\sqrt{3}$; however, $\Omega_{\text{Er},l}$ is not restricted by the dipole interaction strength and could be made larger than $\Omega$ to further decrease the total gate time and increase fidelity. The solution $F_\text{CNOT}$ to first-order in $\gamma_{k,l},\gamma^\star_{k},\chi_k\ll \Omega\propto\Delta\nu$ for all $k,l$ and second-order in $\epsilon\ll \pi$ and $\xi=\delta\nu/\Delta\nu\ll1$ is
    \begin{equation}
    \label{CNOTgateF}
   		F_\text{CNOT} \simeq 1-T_\text{CNOT}\Gamma-\epsilon^2-\frac{13\pi}{16}\epsilon\xi-\frac{43\pi^2}{128}\xi^2
    \end{equation}
Here $T_\text{CNOT}=5\pi/\Omega=5\pi\sqrt{3}/\Delta\nu$ is the total gate time and $\Gamma$ is the effective dissipation rate:
	\begin{equation}
    \label{gammaCnot}
		\Gamma\simeq\frac{1}{80}\left[31\gamma_\text{Er}+17\gamma^\star_\text{Er}+8\chi_\text{Er}+11\gamma_\text{Eu}+8\gamma^\star_\text{Eu}+17\chi_\text{Eu}\right],
	\end{equation}
where we define $\gamma_k=\gamma_{k\uparrow}+\gamma_{k\downarrow}$. To obtain the reverse CNOT fidelity, it is only necessary to swap the dissipative parameters for Er$^{3+}$  and $^{151}$Eu$^{3+}$.
    
    
	For a dipole interaction strength of $\Delta\nu=2\pi\times 46$ kHz corresponding to an Er-Eu separation of about 6 nm, the CNOT gate time can be as small as $T_\text{CNOT}=94\hspace{1mm}\mu$s. To estimate the fidelity, we use the parameters $\gamma_\text{Er}=2\pi\times 3$ Hz, $\gamma^\star_\text{Er}= 2\pi\times 8$ Hz, $\gamma_\text{Eu}=2\pi\times 1.3$ Hz, and $\gamma^\star_\text{Eu}= 2\pi\times 19$ Hz estimated from Ref. \cite{mcauslan2009strong}; also $\chi_\text{Er}\simeq 2\pi\times 80$ Hz and $\chi_\text{Eu}\simeq 0$. In this case, for a small $\pi$-pulse over-rotation of $\epsilon\simeq\pi/64$ and a dipole interaction strength over-estimation of $2\%$ ($\xi\simeq 0.02$), the fidelity is $F_\text{CNOT}=0.986$. The reverse CNOT fidelity is slightly smaller due to the spin dephasing of Er: $F_\text{R-CNOT}=0.980$.
	
	This method can also be used to compute the fidelity of state transfer from Er$^{3+}$  to $^{151}$Eu$^{3+}$. In this case, we remove the fourth pulse ($\mathcal{R}_{\text{Eu},\uparrow}$) and use the initial state $\ket{\psi}=(\ket{\uparrow}+\ket{\downarrow})_\text{Er}\ket{\uparrow}_\text{Eu}$. The expected final state is $\ket{\psi_f}=\ket{\downarrow\downarrow}+e^{i2\varphi}\ket{\uparrow\uparrow}$. In this case, the fidelity is similar to the CNOT gate (Eq. (\ref{CNOTgateF})):
    \begin{equation}
    F_\text{ST} \simeq 1-T_\text{ST}\Gamma-\frac{5}{8}\epsilon^2-\frac{3\pi}{16}\epsilon\xi-\frac{21\pi^2}{256}\xi^2
    \end{equation}
    where $T_\text{ST}=4\pi/\Omega=4\pi\sqrt{3}/\Delta\nu$. The effective dissipation rate was found to be very similar to Eq. (\ref{gammaCnot}), but depends less strongly on $^{151}$Eu$^{3+}$ dissipation:
	\begin{equation}
		\Gamma \simeq \frac{1}{80}\left[29\gamma_\text{Er}+16\gamma^\star_\text{Er}+9\chi_\text{Er}+11\gamma_\text{Eu}+7\gamma^\star_\text{Eu}+9\chi_\text{Eu}\right].
	\end{equation}
	Using the same parameters as above, the fidelity of state transfer is $F_\text{ST}=0.989$.
    
	\subsection{State measurement} \label{statemeasurement}
    
    The spin readout of Er$^{3+}$  is performed by optical excitation. Since only the $\ket{\uparrow} - \ket{e_2}$ transition is coupled to the cavity, optical excitation will result in a presence or absence of a photon emission depending on the state of the Er$^{3+}$ ion. The measurement of the $^{151}$Eu$^{3+}$ ion can be performed through its nearby Er$^{3+}$ ion as a readout ion, provided that the Er$^{3+}$ ion is initially prepared in the $\ket{\uparrow}$ state. Performing spin readout of the $^{151}$Eu$^{3+}$ ion in the Z basis is then achieved by optically exciting the $^{151}$Eu$^{3+}$ ion  ($\ket{\uparrow}$ to $\ket{e}$ transition) followed by exciting the Er$^{3+}$ ion ($\ket{\uparrow}$ to $\ket{e_1}$ transition). The Er$^{3+}$ ion will excite to $\ket{e_1}$ (remain in ($\ket{\uparrow}$) if the state of the $^{151}$Eu$^{3+}$ ion is $\ket{\downarrow}$ ($\ket{\uparrow}$) due to the permanent electric dipole-dipole interaction which shifts the Er$^{3+}$  optical resonance. To readout in the X basis, a $\frac{\pi}{2}$ MW pulse should be applied to the $^{151}$Eu$^{3+}$ ion in order to rotate the ground-state spins before the optical excitation step.
	
	State measurement of each Er$^{3+}$ ion requires the detection of an emitted photon.
	Due to the Purcell effect, as discussed in Sec. \ref{cavitysection}, the emission of a single photon from the ion is highly preferential into the cavity mode. For a high quality cavity, the probability that the photon emits into the cavity tends to unity. Therefore, the detection probability will be limited by coupling losses and single photon detectors, which can have detection efficiency as high as $95\%$, as has been demonstrated in superconducting detectors \cite{lita, Nam}. To do better than this limit, it is necessary to pump the Er$^{3+}$ ion into a cycling transition such that many photons will be emitted by the cavity, and eventually detected. Using such a cycling transition, detection probability can be as high as $98.7\%$ \cite{Obrien}.
	The detection efficiency is not $100\%$ in this case because there is a small chance for the ion to decay into a different state than the initial state, thus ending the photon cycling \cite{sellars-cycling}.
	This chance grows linearly with the number of cycles before detection.
    
	\subsection{Spectrally-multiplexed implementation} \label{ssec:multiplexed} 
	Our spectral multiplexing scheme relies on the possibility that many spectral channels can be operated in parallel. This requires that, in one node, different cavities emit photons that feature different carrier frequencies. This can be accomplished by frequency translation. Noise-free translation over tens of gigahertz  \cite{Neil,marcel.li} can be achieved by using voltage-swept, commercially-available, waveguide electro-optic modulators that can be optically-coupled to the output port of each cavity. After frequency translation, the output can be coupled to a common spatial mode (e.g. a waveguide or fiber) by using a tunable ring resonator filter that features resonance linewidths as narrow as 1 MHz \cite{vahala}. Arrayed waveguides or fiber-Bragg gratings may also be used, however they are bulky and their resonance linewidths are currently not at the MHz level. The modulators and filters may be fabricated on a single chip, offering the possibility of low loss and up to 10$^4$ spectral modes. The Bell-state measurement station consists of a beamsplitter, two sets of ring resonator filters which are identical to that used at the nodes, and an array of superconducting nanowire-based photon detectors, chosen due to their combination of high-efficiency and low noise properties \cite{Nam}. 
	
	\section{Conclusion}\label{Sec:Conclusions}
	
Our proposal for a quantum repeater which is based on individual RE ions promises the deterministic establishment of high-fidelity entanglement over long-distances at a rate which exceeds that corresponding to the direct transmission of photons. Our scheme utilizes some of the most desirable features of RE-ion-doped crystals, specifically emission within the low-loss telecommunications window (Er$^{3+}$) and the hours-long nuclear spin coherence lifetime ($^{151}$Eu$^{3+}$:Y$_2$SiO$_5$) that is needed to perform long-distance transmission and swapping of entanglement. Moreover, control logic gates between close-lying individual $^{151}$Eu$^{3+}$ and Er$^{3+}$ ions allow the quasi-deterministic swapping of entanglement by means of a permanent electric dipole-dipole interaction. The multiplexed version of our scheme improves the entanglement distribution rate by at least a factor of 100 over that of the single-mode version of our repeater.

Looking forward, it is interesting to consider the possibility of employing individual $^{167}$Er$^{3+}$ ions instead of Er-Eu ion pairs for a telecommunication wavelength quantum repeater. In the presence of strong magnetic fields, $^{167}$Er$^{3+}$:Y$_2$SiO$_5$ features a nuclear spin coherence lifetime in the one-second range \cite{Er167}, allowing the possibility of entanglement generation and storage using the same ion, pairs (or small ensembles) of Er$^{3+}$ ions. One of the main challenges for future work in this direction is to devise a scheme whereby individual Er$^{3+}$ ions may be addressed and coupled within a single cavity. This could be achieved by using magnetic dipole-dipole interactions in a similar spirit to what has been demonstrated using nitrogen vacancy centers and carbon spins in diamond \cite{NV-Carbon}, or by using the cavity mode to mediate the interaction. Another interesting direction is the possibility of long-term storage using host-ion spins such as yttrium in Y$_2$SiO$_5$ \cite{Y}.

	\section*{ACKNOWLEDGMENTS}
 The authors would like to thank M. Afzelius, P. Barclay, P. A. Bushev, C. W. Thiel and W. Tittel for helpful discussions. This work was supported by the Natural Sciences and Engineering Research Council (NSERC) through its Discovery Grant program, the CREATE grant 'Quanta', and through graduate scholarships, by the University of Calgary through an Eyes High postdoctoral fellowship, by Alberta Innovates Technology Futures (AITF) through graduate scholarships, and by the Defense Advanced Research
Projects Agency (DARPA) through the Quiness program (Contract No. W31P4Q-13-1-0004).

	\bibliographystyle{apsrev4-1}
	\bibliography{ref}

	\renewcommand{\theequation}{SI\arabic{equation}}
	
	\setcounter{table}{0}
	\setcounter{equation}{0}
	\setcounter{figure}{0}
	\makeatletter
	\renewcommand{\bibnumfmt}[1]{[SI#1]}
	\renewcommand{\citenumfont}[1]{SI#1}

	\widetext

\end{document}